# Efficient transverse multi-wave interactions up to six-wave mixing in a high-Q lithium niobate microresonator

Chuntao Li[1,5,8], Ni Yao[2,8], Huakang Yu[3,8], Jintian Lin[4,*], Renhong Gao[5], Jiale Deng[3], Jianglin Guan[1,5], Lingling Qiao[4], and Ya Cheng[1,4,5,6,7,†]

[1]*State Key Laboratory of Precision Spectroscopy, East China Normal University, Shanghai 200062, China*
[2]*Research Center for Frontier Fundamental Studies, Zhejiang Lab, Hangzhou, 311100, China*
[3]*School of Physics and Optoelectronics, State Key Laboratory of Luminescent Materials and Devices, South China University of Technology, Guangzhou 510460, China*
[4]*State Key Laboratory of High Field Laser Physics and CAS Center for Excellence in Ultra-Intense Laser Science, Shanghai Institute of Optics and Fine Mechanics (SIOM), Chinese Academy of Sciences (CAS), Shanghai 201800, China*
[5]*The Extreme Optoelectromechanics Laboratory (XXL), School of Physics and Electronic Science, East China Normal University, Shanghai 200241, China*
[6]*Shanghai Research Center for Quantum Sciences, Shanghai 201315, China*
[7]*Hefei National Laboratory, Hefei 230088, China*
[8]*These authors contributed equally to this work.*
*jintianlin@siom.ac.cn
†ya.cheng@siom.ac.cn

Feb. 11, 2025

**Abstract**

High-order nonlinear optical processes beyond four-wave mixing ($\chi^{(3)}$) are crucial for advancing ultraviolet (UV) light sources and quantum technologies, yet their practical implementation remains challenging due to inherently weak high-order nonlinear susceptibilities and stringent broadband phase-matching requirements – limitations that persist even in state-of-the-art high-Q microresonators. A breakthrough transverse multi-wave mixing scheme is proposed in a high-Q lithium niobate microresonator, only under single continuous-wave (CW) laser pump. Our approach leverages self-organized subwavelength photorefractive gratings (SPGs) generated through bidirectional stimulated Raman scattering (SRS) process in the microresonator, without using two external counterpropagating lasers. Under single-wavelength pumping (1546 nm), bidirectional SRS (1713 nm) spontaneously generates dynamic SPGs that enable dual-function control: (1) broadband momentum compensation ($\Delta k \approx 2\pi/\Lambda$, $\Lambda$ = grating period) to resolve broadband phase-matching challenges in high-order nonlinear interactions, and (2) preservation of ultrahigh quality factor ($Q > 7 \times 10^6$) for enhanced nonlinear conversion. Moreover, unlike conventional longitudinal configurations requiring more rigid multi-wavelength resonance alignment, our transverse architecture decouples nonlinear processes from cavity-mode constraints through SPG-mediated momentum engineering, enabling simultaneous support for sum-frequency generation to six-wave mixing processes across 500 nm bandwidth with high conversion efficiencies. Furthermore, cascaded SRS process is simultaneously activated to generate light signal for subsequent nonlinear interactions. Consequently, this novel approach enables the first demonstration of single-pump phase-matched transverse SFG with record conversion efficiency (590%/W). And transverse multi-wave mixing processes from $\chi^{(3)}$ (four-wave) to $\chi^{(5)}$ (six-wave) processes are achieved the first time using only the single CW pump, leading to efficient visible and UV light generation, representing a notable advance in nonlinear integration.

Nonlinear frequency mixing in integrated photonics has emerged as a cornerstone technology for developing reconfigurable optical sources [1-5], with particular emphasis on exploiting high-order nonlinear effects beyond conventional four-wave mixing ($\chi^{(3)}$). The exploration of the exploration of high-order nonlinear phenomena beyond the $\chi^{(4)}$ regime [6] holds transformative potential for quantum photonics [7,8], ultrafast optical systems [9], and ultraviolet (UV) light generation [10,11] – critical domains for next-generation manufacturing and metrological applications.

However, practical implementation on chip-scale platforms remains fundamentally challenging. The inherent exponential decay of nonlinear susceptibility with increasing order severely limits nonlinear conversion efficiency under conventional excitation schemes. Even in high-quality-factor ($Q > 10^6$) microresonators that dramatically enhance light-matter interaction, practical implementation of multi-wave mixing processes involving five or more photons requires simultaneous satisfaction of two stringent conditions: (1) broadband phase matching across multiple octaves and (2) multi-wavelength resonance alignment within a single spatial mode family. This dual requirement creates a multidimensional optimization problem that current dispersion engineering techniques struggle to resolve, particularly when targeting visible-to-UV spectral regions where material dispersion becomes dominant.

In this work, we demonstrate a paradigm-shifting approach enabling efficient transverse nonlinear interactions up to six-photon processes ($\chi^{(5)}$) in a single lithium niobate microresonator, using single-wavelength pumping at telecom band (1546 nm). The breakthrough stems from the first combination of transverse nonlinear configuration [12,13] and self-organized subwavelength photorefractive gratings (SPGs), enabled by the outstanding optical property of lithium niobate on insulator [14-16]. These SPGs are spontaneously generated through bidirectional stimulated Raman scattering (SRS) in the microresonator at 1713 nm, eliminating the need for dual counterpropagating pumps [17]. These dynamically reconfigured SPGs provide 1) broadband momentum compensation ($\Delta k \approx 2\pi/\Lambda$, $\Lambda$: grating period) and 2) preserve ultrahigh Q-factor ($>7\times10^6$), enabling unprecedented conversion efficiencies while relaxing stringent resonance requirements. Moreover, distinct from conventional longitudinal schemes that require much stricter multi-wavelength resonance alignment, our transverse configuration decouples nonlinear processes from cavity mode constraints through SPG-mediated momentum engineering. Furthermore, a cascaded SRS process at 1922 nm further initiates subsequent nonlinear interactions. As a result, this innovation enables broadband phase matching across 500 nm while maintaining high conversion efficiency for all nonlinear orders from second-order ($\chi^{(2)}$) to $\chi^{(5)}$. Consequently, single-pump phase-matched transverse sum-frequency generation (SFG) is achieved with record conversion efficiency, and efficient UV light generation is even demonstrated at 350 nm. The demonstrated platform establishes new possibilities for chip-scale nonlinear photonics, particularly for multi-octave frequency comb generation and wavelength-agile UV source development.

**Results**

**Transverse multi-wave mixing processes up to six-wave mixing process ($\chi^{(5)}$)**

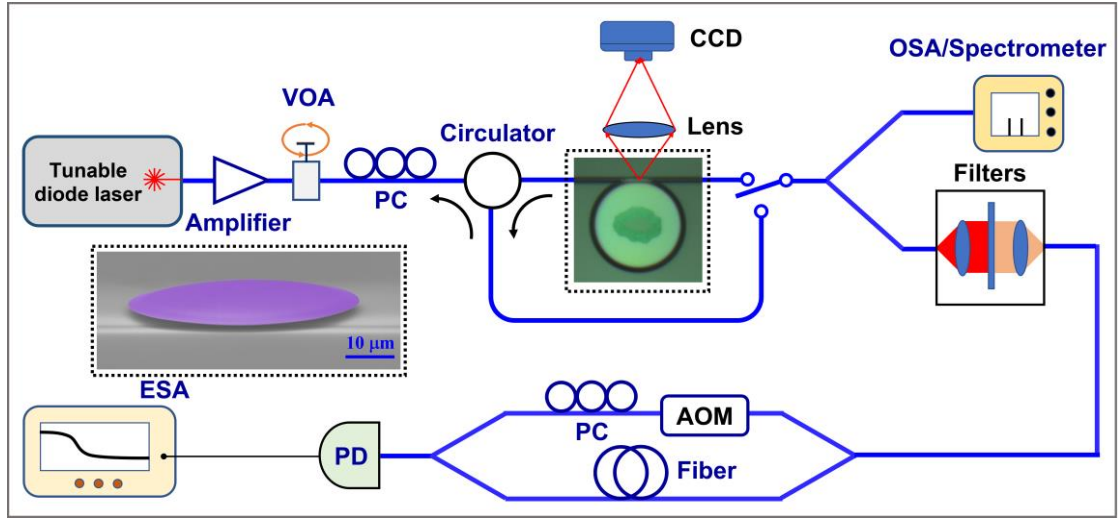

FIG. 1 Experimental setup. Inset: Experimental setup for short-wavelength polarization analysis and power measurement. Here, variable optical attenuator, polarization controller, optical spectral analyzer, acousto-optic modulator, electrical spectrum analyzer, and photo-electric detector, are denoted as VOA, PC, OSA, AOM, ESA, and PD, respectively. Inset: Scanning-electron-microscope image of the microdisk.

The X-cut microdisk employed for generating transverse multi-wavelength mixing processes is depicted in the inset of Fig.1. The diameter, thickness, and wedge angle of the sidewall were ~44.9 μm, ~0.6 μm, and ~17°, respectively (see Supplementary S1 for the details of microdisk fabrication) [18,19]. The experimental setup for the nonlinear interaction is schematically illustrated in Fig. 1. A CW laser from a 1550-nm tunable diode laser, amplified by an erbium-doped fiber amplifier, serves as the transverse-electrically (TE) polarized pump light. This pump light is coupled into the microdisk through a tapered fiber. Besides the forward SRS and cascaded SRS signals generated in the microdisk, the backward SRS and cascaded SRS signals are also collected for spectral analysis and power measurement (see Supplementary S2 for further details [20]). The loaded Q factor of the pump mode is measured to be as high as $7.43\times10^6$, facilitating significant cavity built-up enhancement of the transverse nonlinear processes and modal broadening for mode-hop-free pump tunning (refer to Supplementary S3 for additional details [21-23]).

The forward SRS signal (R-I1) is detected at $\lambda_{R-I1}$ ~1713.2 nm wavelength, corresponding to the Raman shift of 632 cm$^{-1}$, when the pump wavelength is 1545.7 nm with a pump level ≥ 0.45 mW. When further increasing the pump power to ≥ 0.50 mW, forward cascaded SRS signal (R-I2) at $\lambda_{R-I2}$ ~1921.8 nm is detected, as depicted in Fig. 2(a). And the backward SRS signal and cascaded SRS signal also detected with almost the same output powers as the forward counterparts, as shown in Fig. 2(b). Meanwhile, strong backward pump signal also appears in the spectrum, thanks to the slightly deformed microresonator and the significant cavity built-up enhancement. Figure 2(c) plots the output powers of these two forward Raman peaks at different pump powers, both showing a linear relationship. By linearly fitting the output powers of these Raman signals, the extracted conversion efficiencies of these signals are 46.8% and 26.5%, respectively. Moreover, we find that the pump threshold values of these two Raman signals are 0.45 mW and 0.50 mW, respectively. Such low pump threshold values are contributed from the ultra-high Q modes resonant with the pump light and the generated Raman peaks. Specially, these strong forward and backward SRS signals at 1713.2 nm serve as two counterpropagating light waves in the microresonator. It is worth noting that previous work has unveiled that SPGs can be formed in LN microresonators resulting from strong photorefractive effect, when two external counterpropagating pump lasers are coupled into the microresonators with sub-mW powers [17]. Therefore, these reconfigurable SPGs can be spontaneously formed via the bidirectional SRS signals here, as schematically illustrated in the inset of Fig. 3(e), without using two external lasers, which notably reduce the experimental complicacy and provide additional momentum for phase matching. And the intrinsic linewidth of the SRS signal is measured to be 7.8 kHz, as shown in Fig. 2(d).

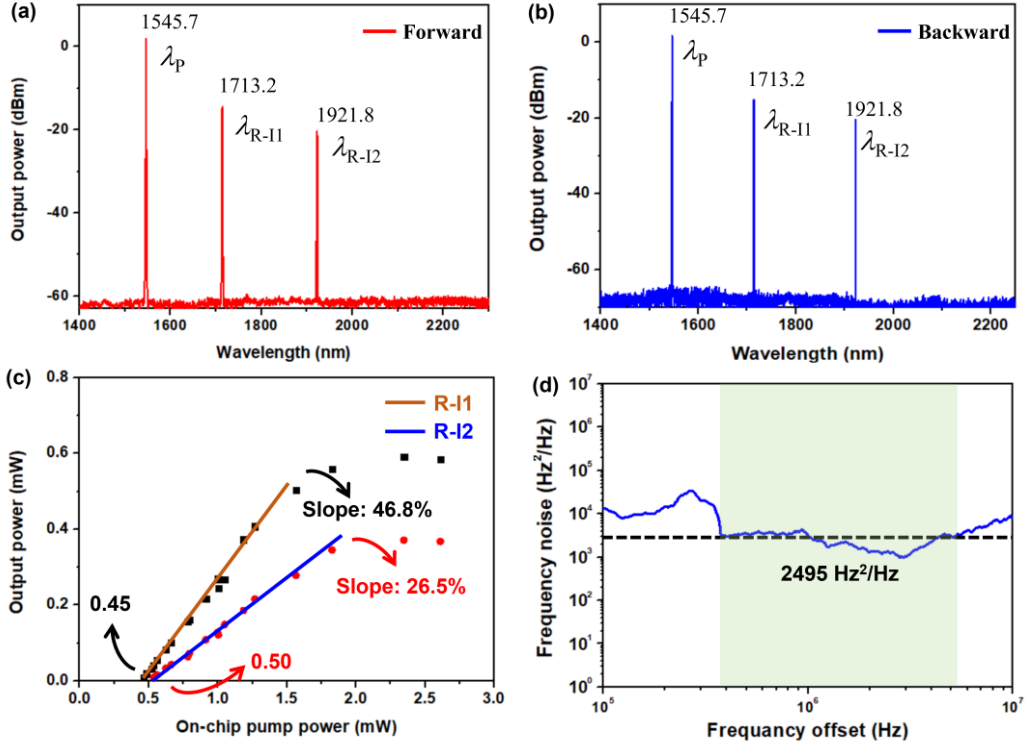

FIG. 2 SRS and cascaded SRS processes. (a) Spectrum of the forward SRS signal (R-I1) and the cascaded SRS signal (R-I2). (b) Spectrum of the backward SRS signal (R-I1) and the cascaded SRS signal (R-I2). (c) Output power of the two forward Raman peaks varied with the pump power. (d) Frequency noise spectrum of the signal measured by the delayed self-heterodyne measurement. A white noise floor is highlighted by the shadow box.

When the pump wavelength is tuned to 1545.77 nm at 0.8~3.9 mW pump level, a nonlinear signal (TSFG) is generated at ~856.7 nm, as shown in Fig. 3(b). According to energy conservation law, this signal is generated by SFG by mixing the pump light and the cascaded SRS signal R-I2. Optical light emitted from the microdisk is recorded, as shown in the sub-plot TSFG of Fig. 3(a). Notably, there are fine glow fringe microstructures periodically distributed around the resonator circumference, showing a period ~ 3.3 µm. Moreover, the total periodic fringe profile is anisotropic, resulting from the strong birefringence, which breaks the rotational symmetry, for the X-cut LN microdisk. Thus, the optical field will deviate from the symmetric one [24,25]. The responding output power of the signal TSFG reaches ~ -10 dBm, as shown in Fig. 3(b). The output power of the signal TSFG grows linearly with the square of the increasing on-chip pump power, showing a conversion efficiency of 590%/W, as illustrated in the inset of Fig. 3(b), and agreeing well with the nature of SFG process.

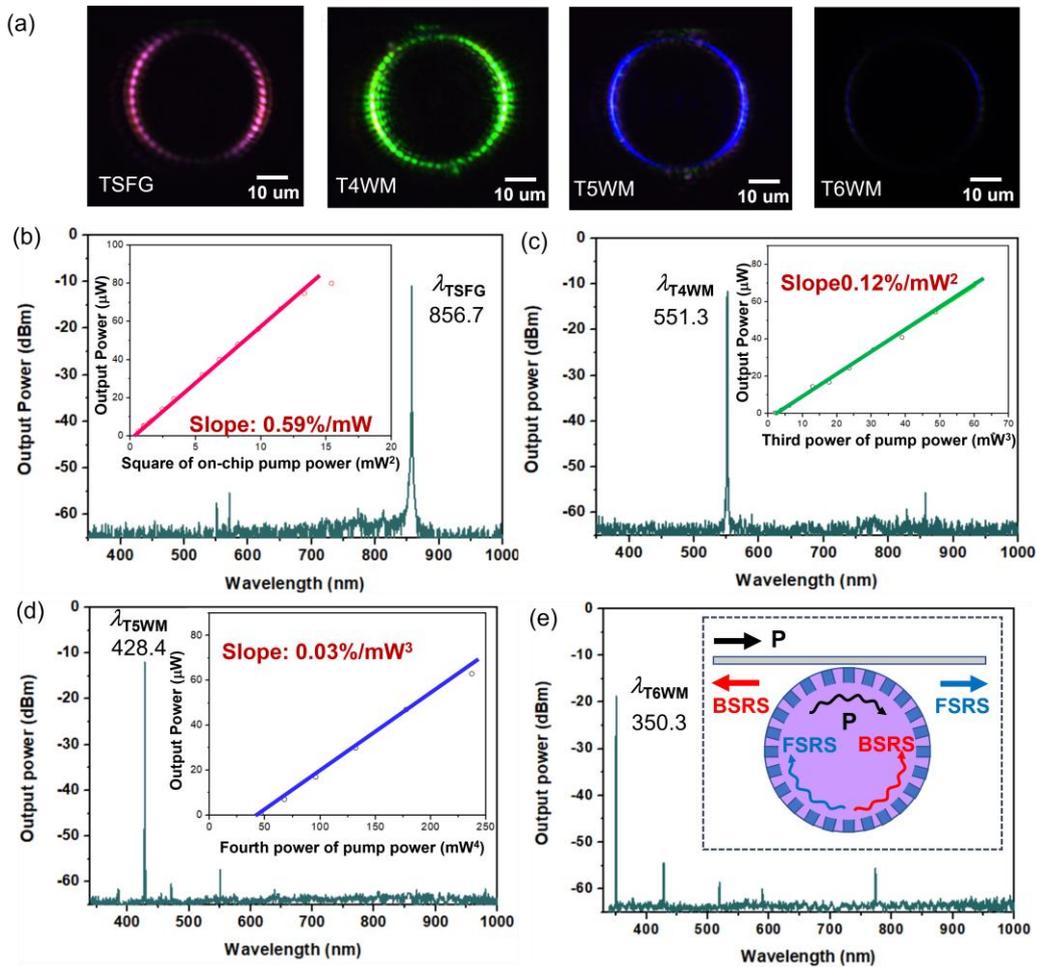

FIG. 3 Multi-wave mixing processes. (a) Observation of light emissions corresponding to the different nonlinear processes. (b) Spectrum of the SFG signal TSFG. Inset: output power of the signal as a function of pump power. (c) Spectrum of the four-wave mixing signal T4WM. Inset: output power of the signal as a function of pump power. (d) Spectrum of the five-wave mixing signal T5WM. Inset: output power of the signal as a function of pump power. (e) Spectrum of the six-wave mixing signal T6WM. Inset: Schematic of the SPGs formed by bidirectional SRS signals in the microresonator.

When the pump wavelength is further tuned to 1545.86 nm at 1.2~3.9 mW pump level, a new signal (T4WM) is generated at ~551.3 nm wavelength. The captured optical emission is provided in the sub-plot T4WM of Fig. 3(a), exhibiting a green fringe period ~ 3.3 µm. And the signal TSFG signal almost vanishes, because of the deviation from the optimal phase-matched condition as the pump wavelength red shifts. According to energy conservation law, the T4WM photon is generated through the four-wave mixing process via mixing two pump photons and one R-I2 photon. The responding output power of the signal T4WM is ~ -11 dBm, as shown in Fig. 3(c). The signal T4WM linearly grows with the third power of the pump power, as shown in the inset of Fig. 3(c), and the conversion efficiency is measured to $0.12\%/mW^2$.

When the pump wavelength is tuned to 1546.00 nm with on-chip pump power of 2.6~3.9 mW, a new signal (T5WM) with bright blue emission is detected at 428.4 nm (see optical emission from the microresonator in the sub-plot T5WM of Fig. 3(a)), which is confirmed by the spectrum shown in Fig. 3(d). And there is a big detuning Δ of the optimum phase matching condition for the four-wave mixing process as the pump wavelength increases, leading to the disappearance of the signal T4WM. Thus, according to energy conservation law, this new emerged T5WM photon is generated through five-wave mixing process by mixing two pump photons and two R-I2 photons. The responding output power of the signal T5WM reaches ~ -12 dBm, as shown in Fig. 3(d). The inset of Fig. 3(d) shows that the output power of the signal T5WM grows biquadratically with the increase of the on-chip pump power, exhibiting a conversion efficiency of $0.03\%/mW^3$.

When the pump wavelength is even tuned to 1546.06 nm at 3.9 mW pump level, another nonlinear signal (T6WM) appears at 350.3 nm, which is closed to the border of the transparency window 330 nm of LN, as shown in Fig. 3(e). The measured output power of the ultraviolet signal is ~ -19 dBm. And the signal T5WM is extinct remarkable. This new emerged photon is ascribed to the six-wave mixing process by mixing two pump photons and three R-I2 photons, representing the highest order multi-wave mixing process demonstrated in microcavity so far [6]. Since the detuning of optimum phase matching condition for the five-wave mixing process is relatively small, there is a weak residual blue signal, which is confirmed by the captured optical emission shown in the sub-plot T6WM of Fig. 3(a).

**Phase matching analysis**

For laser beams with same frequency and wavevector are coupled into the optical waveguide or cavity in a counterpropagating configuration assisted with SPGs, it is known that transverse second harmonic generation (TSHG) is detected in the directions perpendicular to the optical waveguide or cavity, as required by wave-vector matching [12,13,25]. Similarly, as depicted in Fig. 4, for counterpropagating laser beams with different frequencies or wavevectors, transverse sum frequency generation (TSFG) with tilted angle $\theta$ could be expected, where the angle $\theta$ meets the following relationship with wavevector difference $\Delta\beta$, between the two interacting waves and the wave vector $\Lambda = 2\beta_{R-I1}$ of the SPGs,

$$\sin\theta = \frac{\Delta\beta}{k_{SFG}} = \frac{\Delta\beta}{2\pi/\lambda_{SFG}} \quad (1)$$

If there are another two counterpropagating laser beams as depicted in Figs. 4(a)-(c), then we could expect two outgoing beams of TSFG with intersection angle of $2\theta$.

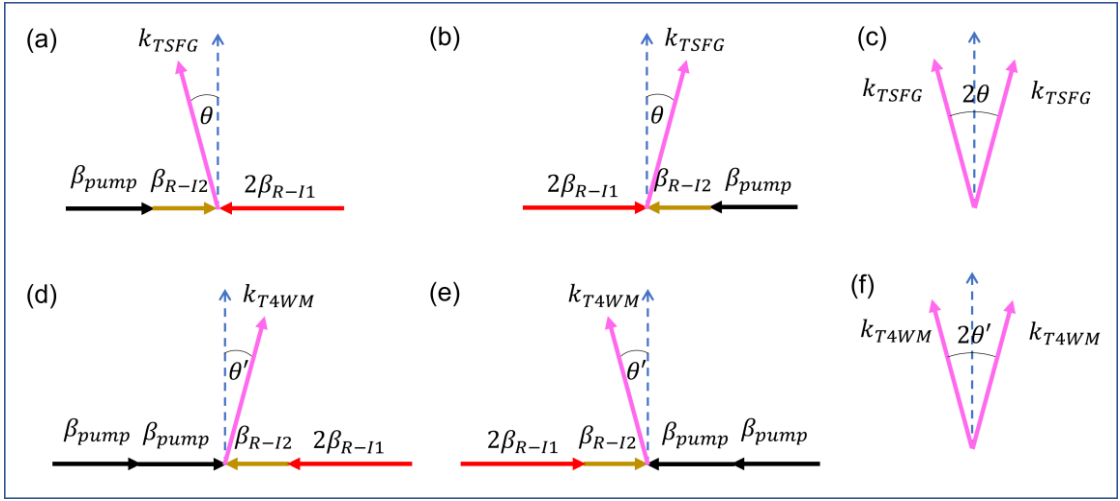

FIG. 4 (a)-(c) Wavevector diagram of phase-matching condition for TSFG. (d)-(f) Wavevector diagram of phase-matching condition for transverse four-wave mixing.

Recalling the principle of two laser beam interference, the period $\Lambda'$ of interference fringes could be determined as follows if the intersection angle of the two laser beams is $2\theta$,

$$\Lambda' = \frac{\lambda}{2 \sin 2\theta} \quad (2)$$

Notably, there is clear red glow fringes observed from the LN microcavity, as shown in Fig. 3(a), representing the coherent built-up lengths of longitudinal SFG. These red glow fringes are believed as TSFG from a pump cavity mode ($\lambda_P$ =1546 nm) and a cascaded Raman mode ($\lambda_{R-I2}$ =1922 nm), as the wavelength of the observed TSFG (856.7 nm) is precisely the SFG of the two mentioned modes. We could readily determine the fringe period ~ 3.3 µm from Fig. 3(a). Using Eq. (S2), the intersection angle $2\theta$ can be estimated to be 7.03°. Now one could easily obtain the wavevector difference between the two interacting cavity modes $\Delta\beta_{exp} = 4.5 \times 10^5 m^{-1}$.

Alternatively, it is known that the wavevector difference between TFSG assisted with SPGs in resonator can be expressed as,

$$\Delta\beta = \frac{m_{pump} + m_{R-I2} - 2m_{R-I1}}{R}, \qquad (3)$$

where $\Delta\beta$ is dependent on the azimuthal number difference of two interacting optical modes and the SPGs (i.e., pump, cascaded SRS, and SRS modes, respectively) and resonator radius $R$. Considering the outer radius of the cavity was ~ 22.45 μm, we could then have $m_{pump} + m_{R-I2} - 2m_{R-I1} = 11$. This azimuthal number difference indicates an occurrence of intermodal Raman scattering. These corresponding modes for the pump, SRS, and cascaded SRS modes are attributed to whispering gallery modes $TE_{0,160}$, $TE_{4,120}$, and $TE_{7,91}$, respectively. The subscript ($n$, $m$) of the mode $TE_{n,m}$ are radial and azimuthal mode numbers, respectively. Therefore, the SPGs provide additional momentum to enable phase-matched transverse SFG process.

As to the generation of the signal T4WM at 551.3 nm, the green glow fringes are also observed with period of ~3.3 μm, as shown in Fig. 3(a). Therefore, the azimuthal number difference is inferred to be 11, agreeing well with the calculated results ( $2m_{pump} - m_{R-I2} - 2m_{R-I1} = -11$ ). Furthermore, the detailed phase-matched analysis for the higher-order nonlinear interactions can be found in the Supplementary S4 [26]. As a result, the SPGs provide additional momentum to enable broadband phase-matched transverse multi-wave mixing processes.

In summary, by leveraging self-organized SPGs formed in the high-Q LN microresonator, we have demonstrated efficient transverse nonlinear interactions up to six-wave mixing under single CW pump laser pump, for the first time to the best of our knowledge. This platform establishes a universal framework for efficient high-order nonlinear photonics, offering transformative potential for applications ranging from ultra-low-threshold frequency conversion, miniaturized optical instrument to multi-channel quantum light sources.

We thank Prof. Yuan Yao, Dr. Botao Fu at East China Normal University and Prof. Ming Li at University of Science and Technology for helpful discussions on frequency noise analysis, anisotropic microresonator analysis, and multi-wave mixing processes, respectively.

**References:**

[1] D. V. Strekalov, C. Marquardt, A. B. Matsko, H. G. L. Schwefel, and G. Leuchs, J. Opt. **18**, 123002 (2016).

[2] I. Breunig, Laser Photonics Rev. **10**, 569 (2016).

[3] A. W. Elshaari, W. Pernice, K. Srinivasan, O. Benson, and V. Zwiller, Nat. Photon. 14, 285 (2020).

[4] Y. F. Xiao, C. L. Zou, Q. Gong, and L. Yang, *Ultra-high-Q Optical Microcavities* (World Scientific, 2020), pp. 1-403.

[5] H. Jin, F. M. Liu, P. Xu, J. L. Xia, M. L. Zhong, Y. Yuan, J. W. Zhou, Y. X. Gong, W. Wang, and S. N. Zhu, Phys. Rev. Lett. **113**, 103601 (2014).

[6] J.-Q. Wang, Y.-H. Yang, M. Li, H. Zhou, X.-B. Xu, J.-Z. Zhang, C.-H. Dong, G.-C. Guo, and C.-L. Zou, Nat. Commun. **13**, 6223 (2022).

[7] M. A. Ciampini, A. Orieux, S. Paesani, F. Sciarrino, G. Corrielli, A. Crespi, R. Ramponi, R. Osellame, and P. Mataloni, Light Sci. Appl. **5**, e16064 (2016).

[8] J. C. Adcock, C. Vigliar, R. Santagati, J. W. Silverstone, M. G. Thompson, Nat. Commun. **10**, 3528 (2019).

[9] C. Winterfeldt, C. Spielmann, and G. Gerber, Rev. Mod. Phys. **80**, 117 (2008).

[10] C. Zhang, T. Ooi, J. S. Higgins, J. F. Doyle, L. von der Wense, K. Beeks, A. Leitner, G. A. Kazakov, P. Li, P. G. Thirolf, T. Schumm, and J. Ye, Nature **633**, 63 (2024).

[11] S. Zhu, Y. Zhu, and N. Ming, Science **278**, 843 (1997).

[12] Y. J. Ding, S. J. Lee, and J. B. Khurgin, Phys. Rev. Lett. **75**, 429 (1995).

[13] H. Yu, Y. Lun, J. Lin, Y. Li, X. Huang, B. Liu, W. Wu, C. Wang, Y. Cheng, Z. Li, and J. B. Khurgin, Laser Photonics Rev. **17**, 2201017 (2023).

[14] J. Lin, F. Bo, Y. Cheng, and J. Xu, Photon. Res. **8**, 1910 (2020).

[15] Y. Jia, L. Wang, and F. Chen, Appl. Phys. Rev. **8**, 011307 (2021).

[16] Y. Zheng and X. Chen, Adv. Phys. X **6**, 1889402 (2021).

[17] J. Hou, J. Zhu, R. Ma, B. Xue, Y. Zhu, J. Lin, X. Jiang, X. Chen, Y. Cheng, L. Ge, Y. Zheng, and W. Wan, Laser Photonics Rev. **18**, 2301351 (2024).

[18] See Supplemental Material [url] for the fabrication of the high-Q microresonator, which includes Ref. [19].

[19] R. Wu, J. Zhang, N. Yao, W. Fang, L. Qiao, Z. Chai, J. Lin, and Y. Cheng, Opt. Lett. **43**, 4116 (2018).

[20] See Supplemental Material [url] for the experimental setup.

[21] See Supplemental Material [url] for characterization of Q factor and modal thermo-optic broadening, which includes Ref. [23,24].

[22] J. Wang, B. Zhu, Z. Hao, F. Bo, X. Wang, F. Gao, Y. Li, G. Zhang, and J. Xu, Opt. Express **24**, 21869 (2016).

[23] S. Spillane, T. Kippenberg, and K. Vahala, Nature **415**, 621 (2002).

[24] A. Gao, C. Yang, L. Chen, R. Zhang, Q. Luo, W. Wang, Q. Cao, Z. Hao, F. Bo, G. Zhang, and J. Xu, Photon. Res. **10**, 401 (2022).

[25] Y.-H. Yang, X.-B. Xu, J.-Q. Wang, M. Zhang, M. Li, Z.-X. Zhu, Z.-B. Wang, C.-H. Dong, W. Fang, H. Yu, G.-C. Guo, and C.-L. Zou, Phys. Rev. Appl. **19**, 034087 (2023).

[26] See Supplemental Material [url] for analysis of the broadband phase matching condition for transverse five-wave mixing and six-wave mixing.